\documentclass[journal]{IEEEtran}
\usepackage{amsmath,amsfonts}
\usepackage{algorithmic}
\usepackage{algorithm}
\usepackage{array}
\usepackage[caption=false,font=normalsize,labelfont=sf,textfont=sf]{subfig}
\usepackage{textcomp}
\usepackage{stfloats}
\usepackage{multirow}
\usepackage{url}
\usepackage{verbatim}
\usepackage{graphicx}
\usepackage{cite}
\begin{document}

\title{Artificial Neural Networks Trained on Noisy Speech Exhibit the McGurk Effect}

\author{Lukas Grasse
        and Matthew S. Tata
        
\thanks{L. Grasse and M. S. Tata are with the Canadian Centre for Behavioural Neuroscience, Department of Neuroscience, University of Lethbridge, Lethbridge, AB, Canada. (Corresponding author: Lukas Grasse; e-mail: lukas.grasse@uleth.ca).}
}

\maketitle

\begin{abstract}
Humans are able to fuse information from both auditory and visual modalities to help with understanding speech. This is demonstrated through a phenomenon known as the McGurk Effect, during which a listener is presented with incongruent auditory and visual speech that fuse together into the percept of illusory intermediate phonemes. Building on a recent framework that proposes how to address developmental 'why' questions using artificial neural networks, we evaluated a set of recent artificial neural networks trained on audiovisual speech by testing them with audiovisually incongruent words designed to elicit the McGurk effect. We show that networks trained entirely on congruent audiovisual speech nevertheless exhibit the McGurk percept.  We further investigated 'why'  by comparing networks trained on clean speech to those trained on noisy speech, and discovered that training with noisy speech led to a pronounced increase in both visual responses and McGurk responses across all models. Furthermore, we observed that systematically increasing the level of auditory noise during ANN training also increased the amount of audiovisual integration up to a point, but at extreme noise levels, this integration failed to develop. These results suggest that excessive noise exposure during critical periods of audiovisual learning may negatively influence the development of audiovisual speech integration. This work also demonstrates that the McGurk effect reliably emerges untrained from the behaviour of both supervised and unsupervised networks, even networks trained only on congruent speech. This supports the notion that artificial neural networks might be useful models for certain aspects of perception and cognition.
\end{abstract}


\section{Introduction}

The human brain has evolved to fuse stimuli across sensory modalities such as vision, audition, and somatosensation to make use of the extra information available when those inputs are congruent. When processing complex stimuli such as speech, the brain is capable of integrating auditory and visual information to improve intelligibility, which is particularly useful as speech is commonly mixed with noise or other distractors \cite{sumby1954visual}.  However, auditory and visual information can be made artificially incongruent, leading to a compelling demonstration of integration called the McGurk Effect \cite{mcgurk1976hearing}. The McGurk Effect occurs when a listener is presented with an auditory phoneme, such as /ba/, and visual lip movements that correspond to a different phoneme, such as /ga/. In this scenario listeners generally report perceiving the intermediate phoneme /da/, rather than either of the auditory or visual phonemes. The tendency to perceive the McGurk effect has been extensively studied and found to develop early, including children \cite{nath2011neural} and even infants \cite{rosenblum1997mcgurk}. Other research has shown that it is enhanced in older adults \cite{sekiyama2014enhanced}, potentially correlated with age-related decline in auditory processing. 

Congruent visual lip movements have long been known to improve speech intelligibility \cite{sumby1954visual,bernstein2004auditory} in the presence of auditory noise.  It follows that a key factor that influences McGurk perception across development is whether or not auditory noise is mixed with the incongruent stimulus. \cite{hirst2018threshold} found that there is a noise threshold above which the listener is more likely to perceive the McGurk effect. They also found that the threshold decreases across development, with younger children requiring more auditory noise to induce perception of the effect whereas adults and older children required little or no noise. This aligns with previous research showing that there is a developmental shift from relying primarily on auditory cues early in childhood to relying more heavily on visual cues throughout development \cite{diaconescu2013visual}. However, the role of noise \textit{during development} has not been explored.

Although the McGurk effect has been established as a useful tool for studying audiovisual integration of speech, there are important differences in how the brain processes incongruent stimuli, such as the McGurk effect. As described in \cite{alsius2018forty}, these differences include large variability in tendency to perceive the McGurk effect across participants, as well as differences in behavioural and neural signatures  between processing of congruent audiovisual stimuli and McGurk stimuli. Some examples of these differences include increased experimental response times for participants when  perceiving McGurk stimuli relative to congruent stimuli \cite{beauchamp2010fmri}, individual differences in the perceptual experience of McGurk vs congruent stimuli \cite{soto2009deconstructing}, and differences in patterns of activation during neuroimaging studies \cite{nath2012neural}. One possible explanation for this variability is that young humans experience a range of different environments during development.  Since the presence of noise in audiovisual stimuli directly impacts the perception of the McGurk Effect, it is likely that the presence of noise during development also affects how the brain learns to use audiovisual speech.

Given these differences, and the importance of audiovisual integration in general for accurate perception, a better understanding of why the McGurk effect arises during development and the factors that contribute to its emergence is an essential area of research. Asking what conditions lead the brain to learn to integrate auditory and visual information is difficult, because we cannot modify human development retrospectively and non-human animal models of human phonemic and linguistic processing are not possible. A recent alternative approach to asking such questions about brain and cognitive development is through the use of artificial neural networks (ANN) as models to ask specific developmental questions \cite{kanwisher2023using}. The idea is that by training ANNs to solve human tasks under various conditions, and then observing which training conditions give rise to human-like behaviour, we begin to answer the question of why the brain develops the way it does. For example, if a perceptual illusion or effect that humans experience also arises in ANNs under certain conditions, this suggests that those particular conditions might matter also for human development.  Kanwisher and colleagues \cite{kanwisher2023using} have proposed four main dimensions across which ANNs can be modulated: the neural network architecture, the input data given to the network, the objective function or loss, and the learning algorithm used during training. This experimental approach enables researchers to test which elements causally contribute to the development of a given behaviour, by rolling back and modifying evolutionary and developmental processes in a way that is impossible to do with humans. 

Use of ANNs as models for human perception is a good approach to ask questions about how humans might develop the audiovisual integration that leads to the McGurk Effect.  If ANNs exhibit the sort of fusion that gives rise to this effect only under certain conditions, this provides evidence that these same conditions might impact human learning. Early research into ANNs for audiovisual speech perception found that, as in humans, including visual lip movement features with noisy audio leads to improved speech recognition \cite{yuhas1989integration}. A foundational paper in the area of audiovisual speech recognition using deep ANNs was \cite{ngiam2011multimodal}, which found that training ANNs using unsupervised learning leads to networks exhibiting the McGurk effect. This study used an autoencoder architecture with Restricted Boltzmann machine layers, and added Gaussian noise to the training audio which they found improved classification performance on a spoken single-digit recognition task.  However, they did not investigate whether adding noise during training affected the tendency of the model to perceive the McGurk effect. An important recent development in ANNs for audiovisual speech recognition has been the transformer architecture \cite{vaswani2017attention}, which has given rise to ANNs that can recognize entire spoken sentences with human-like performance. Early transformer architectures such as Deep AVSR \cite{afouras2018deep} used supervised learning to train networks to recognize speech using large labeled audiovisual datasets such as the LRS2 dataset \cite{afouras2018deep}. More recent architectures have used the more biologically plausible self-supervised learning approach \cite{shi2022avsr,haliassos2022jointly} or data augmentation \cite{ma2023auto} to train transformer models on massive unlabeled datasets.  Thus, state-of-the-art ANNs exhibit human-like audiovisual performance, are trained in a somewhat biologically plausible manner, and are now useful for systematically investigating the conditions that give rise to the McGurk Effect. 

An important discovery of recent work to develop audiovisual ANN architectures is that the incorporation of auditory noise during training improves speech recognition performance and increases the tendency of the network to fuse auditory and visual information. Relatively less research has directly explored the McGurk effect in ANNs. A set of early papers, \cite{sporea2010modelling,sporea2010distributed}  trained feed-forward ANNs on auditory input vectors containing representations of the voice, manner, and place of articulation features from the IPA alphabet and "randomly generated vectors" for the visual features. In \cite{sporea2010modelling}, they tested whether integrating features in the first hidden layer or second hidden layer led to a stronger McGurk effect and found that later integration produced a stronger effect. \cite{sporea2010distributed} tested an auto-associative memory network. Both these papers had the limitation of using hard-coded IPA auditory features and random vectors for visual input. The paper \cite{ngiam2011multimodal} tested audiovisual integration in auto-encoder networks trained on speech characteristics of the spectrogram and pixels-based visual features of the mouth region of interest. This is more biologically-plausible than \cite{sporea2010distributed}, but a limitation of their research was only training on audiovisual datasets consisting of spoken letters and digits. \cite{gustafsson2014self} tested a self-organizing map architecture on audiovisual recordings of a Swedish speaker saying a set of phonemes. The self-organizing map architecture outputs a winner-take-all map for each of the possible responses, and they found that when presenting incongruent stimuli the highest value in the output map corresponded to the McGurk perception option.  

These previous studies provide a good foundation for computationally modeling audiovisual integration using ANNs, often specifically using the McGurk Effect as a well-understood paradigm. The recent development of large audiovisual speech datasets \cite{nagrani2017voxceleb,afouras2018deep,afouras2018lrs3}, along with architectures that can recognize audiovisual speech to human levels of accuracy \cite{afouras2018deep,shi2022avsr} enable us to test which factors lead to the emergence of human-like audiovisual integration, without having to resort to hard-coding input features or training on smaller datasets that do not consist of developmentally-plausible naturalistic stimuli. The present study takes the first step towards this by evaluating recent transformer networks trained on audiovisual speech using a McGurk dataset consisting of real spoken words. We can now explore whether the McGurk Effect emerges during training with naturalistic speech, and whether or not characteristics of the speech \textit{during training} affect the tendency of the network to perceive the McGurk fusion phenomenon.  We examined how adding auditory noise during training influenced the degree of audiovisual integration and perception of the McGurk effect. We also compared the performance of the ANNs to human participants on the same McGurk dataset. Finally, we tested how systematically increasing the level of noise during training of an audiovisual ANN affected its levels of McGurk perception. We found that training with noisy audio increased the percentages of McGurk and visual responses across all networks. Training an ANN with increasing levels of noise increased the tendency of the network to use visual input, and to map incongruent stimuli to the McGurk response. This tendency was shown using a k-nearest neighbours classifier using the cosine distance metric, and we also found that the relative distance of McGurk stimuli to the auditory and audiovisual representations decreased relative to the visual representation.  However, this role of noise has a limit: at the highest noise levels, audiovisual integration failed to develop.  This result suggests a possible hypothesis about why some individuals do not perceive the McGurk Effect if they experienced high levels of noise during a critical window of audiovisual development.   

\section{Methods}

To evaluate the effect of noise on audiovisual perception during learning, we tested artificial neural networks and human participants on McGurk stimuli \cite{mcgurk1976hearing} using a forced-choice task. First, we compared existing ANNs that were trained on audiovisual speech to test the effect of training on clean speech vs. noisy speech. We also introduced a novel, lightweight audiovisual speech network based on contrastive predictive coding (CPC).  We included this network to systematically vary levels of auditory noise during learning to characterize the range over which noise during training might affect audiovisual integration.  The model architecture and training objective is described below.

\subsection{Stimuli}
We used the set of words from \cite{dekle1992audiovisual} that have been previously shown to elicit the McGurk effect. The list of words is shown in Table \ref{tab:words}. For each of the words in the table, five recordings were made from a male speaker and five from a female speaker. Videos were recorded in an acoustically dampened room in front of a black backdrop using a Logitech webcam (Logitech Brio Ultra HD 4K). Audio from each auditory word pairing was then combined with a randomly selected video from the corresponding visual word pairing. All pairings were selected within speaker. Additionally, five recordings of the audiovisual word corresponding to the fused McGurk option were recorded from each individual for use in the forced-choice k-NN classifier described in Section \ref{knn}.

\begin{table}[!htbp]
\centering
\caption {List of McGurk Stimuli Word Pairings}
\begin{tabular}{lll}
\textbf{Auditory} & \textbf{Visual} & \textbf{Fused} \\ \hline
Bat               & Vet             & Vat            \\
Bet               & Vat             & Vet            \\
Bent              & Vest            & Vent           \\
Boat              & Vow             & Vote           \\
Might             & Die             & Night          \\
Mail              & Deal            & Nail           \\
Mat               & Dead            & Gnat           \\
Moo               & Goo             & New            \\
Met               & Gal             & Net           
\end{tabular}
\label{tab:words}
\end{table}

\subsection{Audiovisual Contrastive Predictive Coding}

\subsubsection{Architecture}
Our self-supervised audiovisual model was based on contrastive predictive coding (CPC) \cite{oord2018representation}. CPC is an approach to self-supervised learning that works across many different modalities such as text, audio, and images. The CPC model architecture consists of 3 components, an encoder network, a context network, and a prediction network. These three models are jointly trained; the encoder network encodes input information into a sequence of representational embeddings, while the context network summarizes the encoded embeddings across time, and the predictor network uses the context to predict future embeddings. The loss function used is InfoNCE \cite{oord2018representation}. This is a contrastive loss that trains the network to reduce the distance between future embedding and predicted embeddings, relative to the distribution of other samples from the training dataset. The original CPC paper used a convolutional encoder with a recurrent neural network as the context network, and a linear predictor network. We replaced the predictor network with a single layer transformer as this was found to improve training performance and stability for speech in \cite{kahn2020libri}. Additionally we replaced the LSTM context network with another single-layer transformer. All transformer layers used absolute position encoding as additional input. 

We combined the auditory encoder with a visual encoder that incorporated lip movement information to create an audiovisual encoder. The audiovisual encoder consisted of a pretrained lip-movement feature model from \cite{afouras2018deep}, and a convolutional layer that combined the visual encoder output with the audio encoder output. The output of this final convolutional layer was then used as the input for the context model. 

\subsubsection{Training}

We trained the AudioVisual CPC model using the LRS3 dataset \cite{afouras2018lrs3} and the InfoNCE contrastive loss. Models were trained for 600k steps using a batch size of 16 and a learning rate of 2e-4. Negative samples for the contrastive loss were taken from other samples in the training batch. Similar to the training approach used in \cite{shi2022learning}, we used modality dropout to drop out the visual or auditory modality with 50\% probability. This previous research has shown that such dropout improves integration of visual information during learning and prevents auditory information from monopolizing the learned audiovisual representation. 

We rendered an hour of babble noise from LRS3 by combining 1 hour of speech from 20 talkers and normalizing the audio. This babble noise was then mixed with speech stimuli during training using a signal-to-noise ratio SNR measured in dB. We trained separate versions of the Audiovisual CPC model using noise levels ranging from 10 dB SNR to -15 dB in increments of 5 dB. 

The pretrained AV-HuBERT models contained separate checkpoints trained on either clean speech or noisy speech, whereas the pre-trained Deep AVSR model was always trained with noise. To compare Deep AVSR with AV-HuBERT we also trained a version of Deep AVSR using clean speech. We trained this model on the LRS2 dataset \cite{afouras2018deep} to ensure consistency for comparison with the pre-trained version that used a noisy training set.

\subsection{K-Nearest Neighbours Classification} \label{knn}
We wanted to compare the likelihood of audiovisual fusion into a McGurk stimulus by each network.  One approach would be to train a linear probe classifier to report the perceived word, however this risks the possibility that the linear classifier itself might learn to fuse information in the embedding space that the unsupervised network did not.  A more interpretable approach to evaluate the amount of audiovisual fusion learned by each of the network embeddings is to use a k-nearest neighbours classifier \cite{fix1985discriminatory} to look at clustering within the embedding space itself.  We used this approach to directly inspect the vector distances in embeddings, rather than train a linear classifier to report which word was "perceived" by the networks. As the distance metric for the k-NN algorithm we used the cosine distance with dynamic time warping \cite{sakoe1978dynamic} to account for the audiovisual speech being potentially misaligned in time. For each k-NN classifier we performed 10-fold cross-validation and calculated the 95\% confidence interval using bootstrapping.

\subsection{Participants}
The McGurk word stimuli dataset was presented to a total of 14 participants (1 male; all right-handed) with an age range of 18-30 (mean ± SD: 20.5 ± 2.98). All participants had normal or corrected-to-normal vision, and reported no hearing impairments. Participants were recruited from an undergraduate course and received course credit for their participation. The study complied with the Declaration of Helsinki and was approved by the Human Subjects Ethics Committee of the University of Alberta. Participants were presented with each of the McGurk stimuli in an audiovisual condition as well as an audio only condition for a total of 180 trials. After each trial the participant was presented with text corresponding to the three word options and performed a forced-choice classification by clicking the word they perceived. The experiment was implemented in Psychopy \cite{peirce2019psychopy2} and presented on a iMac computer with Sennheiser headphones. 

\subsection{Experiment 1: Evaluation of Audiovisual Models On McGurk Forced-Choice Task}

We evaluated a variety of recent artificial neural networks trained on audiovisual speech. The networks are outlined in table \ref{tab:networks}.

The networks were trained using either a self-supervised loss or a supervised loss with Connectionist Temporal Classification (CTC) used for speech recognition. The author-provided implementation of AV-HuBERT \cite{shi2022avsr} contained pre-trained checkpoints for AV-HuBERT trained using self-supervised learning on clean speech, self-supervised learning on noisy speech, and fine-tuned speech recognition on noisy speech. We used an online implementation of Deep AVSR \cite{smeetrs_deep_avsr}. The checkpoints provided for Deep AVSR were all trained on audiovisual speech with noise present, and as such we trained the provided implementation using clean speech for comparison.

\begin{table*}
\centering
\caption{\textbf{Audiovisual Networks.} Deep AVSR \protect\cite{afouras2018deep} was trained to perform speech recognition (SR) using ctc and a supervised loss, whereas AV-HuBERT \protect\cite{shi2022avsr} and our AudioVisual CPC network were trained using self-supervised learning (SSL).}

\begin{tabular}{ | p{0.15\linewidth} | p{0.05\linewidth}| p{0.25\linewidth}| p{0.15\linewidth}| }
\hline
Network                                                                  & Year & Task                                                                                                & Training Noise                                                                                     \\ \hline Deep AVSR                                                             & 2018 & \begin{tabular}[c]{@{}l@{}}Supervised Speech Recognition\\ Curriculum Learning\end{tabular}         & \begin{tabular}[c]{@{}l@{}}Babble,\\ 25\% probability\end{tabular}                                 \\ \hline
AV-HuBERT SSL                                                             & 2022 & \begin{tabular}[c]{@{}l@{}}Self-Supervised Learning\end{tabular} & \begin{tabular}[c]{@{}l@{}}Natural Sounds, \\ Music, Babble, Speech, \\ 25\% probability\end{tabular} \\
\hline
AV-HuBERT SR                                                     & 2022 & \begin{tabular}[c]{@{}l@{}} Speech Recognition Fine-tuning\end{tabular} & \begin{tabular}[c]{@{}l@{}}Natural Sounds, \\ Music, Babble, Speech, \\ 25\% probability\end{tabular} \\ \hline
\begin{tabular}[c]{@{}l@{}}AudioVisual CPC \\ (Our Network)\end{tabular} & 2024 & \begin{tabular}[c]{@{}l@{}}Self-Supervised Learning, \\ Contrastive Predictive Coding\end{tabular}  & \begin{tabular}[c]{@{}l@{}}Babble\\ 100\% probability\end{tabular}                               \\ \hline
\end{tabular}
\label{tab:networks}
\end{table*}

For Experiment 1 we also trained our Audiovisual CPC network in a self-supervised manner on the LRS3 pretraining and training set using the InfoNCE contrastive loss. We trained separate versions using either clean audio or noisy audio with the noise level at 0 dB SNR relative to the talker's speech. Noise was always present with 100\% probability during this learning, as compared to the other networks which generally used a 25\% probability of noise being present during learning. 

\subsection{Experiment 2: Varying Levels of Auditory Noise during Learning}

In Experiment 2 we varied the level of auditory noise during learning using the Audiovisual CPC network. We trained the model using increasing levels of noise from 10 dB SNR up to -15 dB SNR with increments of 5 dB. We tested each of the networks using a k-NN classifier with 10-fold cross-validation in the same manner as experiment 1. We also calculated the average cosine-DTW distance of each incongruent McGurk stimuli embedding to the corresponding congruent auditory, visual, and audiovisual word embeddings to evaluate how the relative distance of the three forced-choice options shift as training noise increased.

\section{Results}
\begin{figure*}[!htbp]
  \centering
    \includegraphics[width=1.0\textwidth]{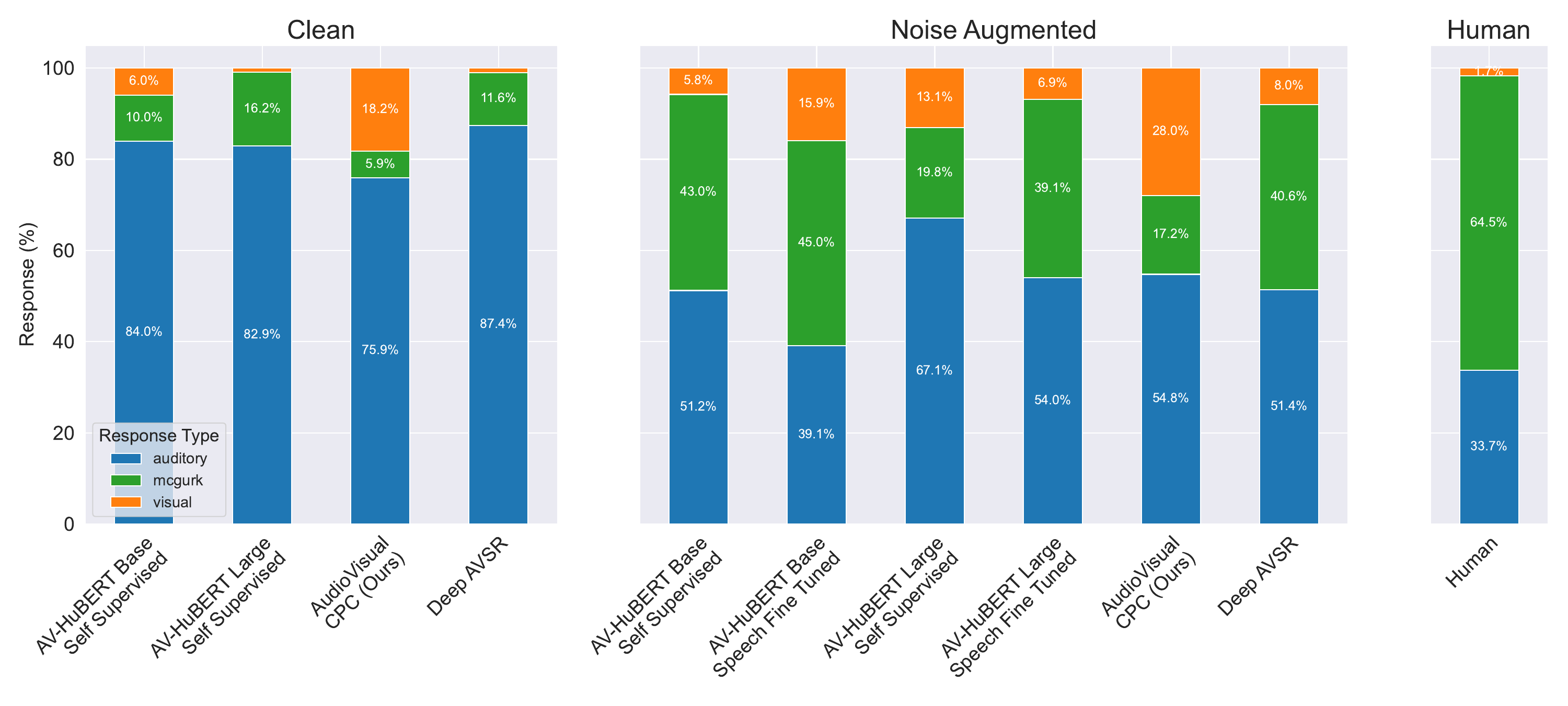}
    \caption{\textbf{McGurk Stimuli Classification Results on Audiovisual Input.} Networks were provided both auditory and visual input, and human participants listened to speech and were presented with a video of the speaker on a computer screen. Networks showed increased McGurk and visual responses when trained with auditory noise relative to their counterparts trained on clean speech. }
    \label{figure:knn_classification_audiovisual}
\end{figure*}

The results of the k-NN classification for each network is shown in Table \ref{tab:cross_val_accuracy}. Most networks could accurately classify the three words in each of the audiovisual pairings. Given this high accuracy on non-McGurk stimuli, it is interesting to note the difference of visual influence on k-NN classifications on the McGurk stimuli. As shown in Figure \ref{figure:knn_classification_audiovisual} all networks saw an increase McGurk and visual classifications when trained on noise-augmented audiovisual speech. A chi-squared test was used to compare each network trained on clean speech with its noise augmented counterparts. The results are as follows:
\begin{itemize}
    \item AV-HuBERT Base (Clean vs. Noisy): $\chi^2(2)=256.47, p=<0.001$
    \item AV-HuBERT Base (Clean vs. Noisy,SFT): $\chi^2(2)=387.97, p=<0.001$
    \item AV-HuBERT Large (Clean vs. Noisy): $\chi^2(2)=114.13, p=<0.001$ 
    \item AV-HuBERT Large (Clean vs. Noisy,SFT): $\chi^2(2)=181.74, p=<0.001$
    \item Deep AVSR (Clean vs. Noisy): $\chi^2(2)=278.23, p=<0.001$ \item Audiovisual CPC (Clean vs. Noisy) : $\chi^2(2)=99.33, p=<0.001$
\end{itemize}
This relative increase is of note, even though the percentage of McGurk responses are still low relative to human participant responses on the same dataset.

\begin{table}[!htbp]
\centering
\caption {\textbf{10-Fold Cross-Validation Accuracy on Congruent Stimuli Training Set.} The shown 95\% confidence intervals are calculated using bootstrapping.}
\begin{tabular}{|l|l|l|l|l}
\cline{1-4}
Network                              & Training Noise & Accuracy & 95\% CI  &  \\ \cline{1-4}
\multirow{2}{*}{Deep AVSR}           & Clean          & 68.52\%  & ± 5.56\% &  \\ \cline{2-4}
                                     & Noisy          & 83.33\%  & ± 4.08\% &  \\ \cline{1-4}
\multirow{2}{*}{AV-Hubert Base SSL}  & Clean          & 94.44\%  & ± 2.41\% &  \\ \cline{2-4}
                                     & Noisy          & 92.96 \% & ± 2.96\% &  \\ \cline{1-4}
AV-Hubert Base SR                    & Noisy          & 98.15\%  & ± 1.48\% &  \\ \cline{1-4}
\multirow{2}{*}{AV-Hubert Large SSL} & Clean          & 95.56\%  & ± 2.41\% &  \\ \cline{2-4}
                                     & Noisy          & 93.33\%  & ± 3.15\% &  \\ \cline{1-4}
AV-Hubert Large SR                   & Noisy          & 92.59\%  & ± 3.52\% &  \\ \cline{1-4}
\multirow{2}{*}{AudioVisual CPC}     & Clean          & 99.26\%  & ± 0.93\% &  \\ \cline{2-4}
                                     & Noisy          & 95.93\%  & ± 2.41\% &  \\ \cline{1-4}
\end{tabular}
\label{tab:cross_val_accuracy}
\end{table}

\begin{figure*}[!htbp]
  \centering
    \includegraphics[width=1.0\textwidth]{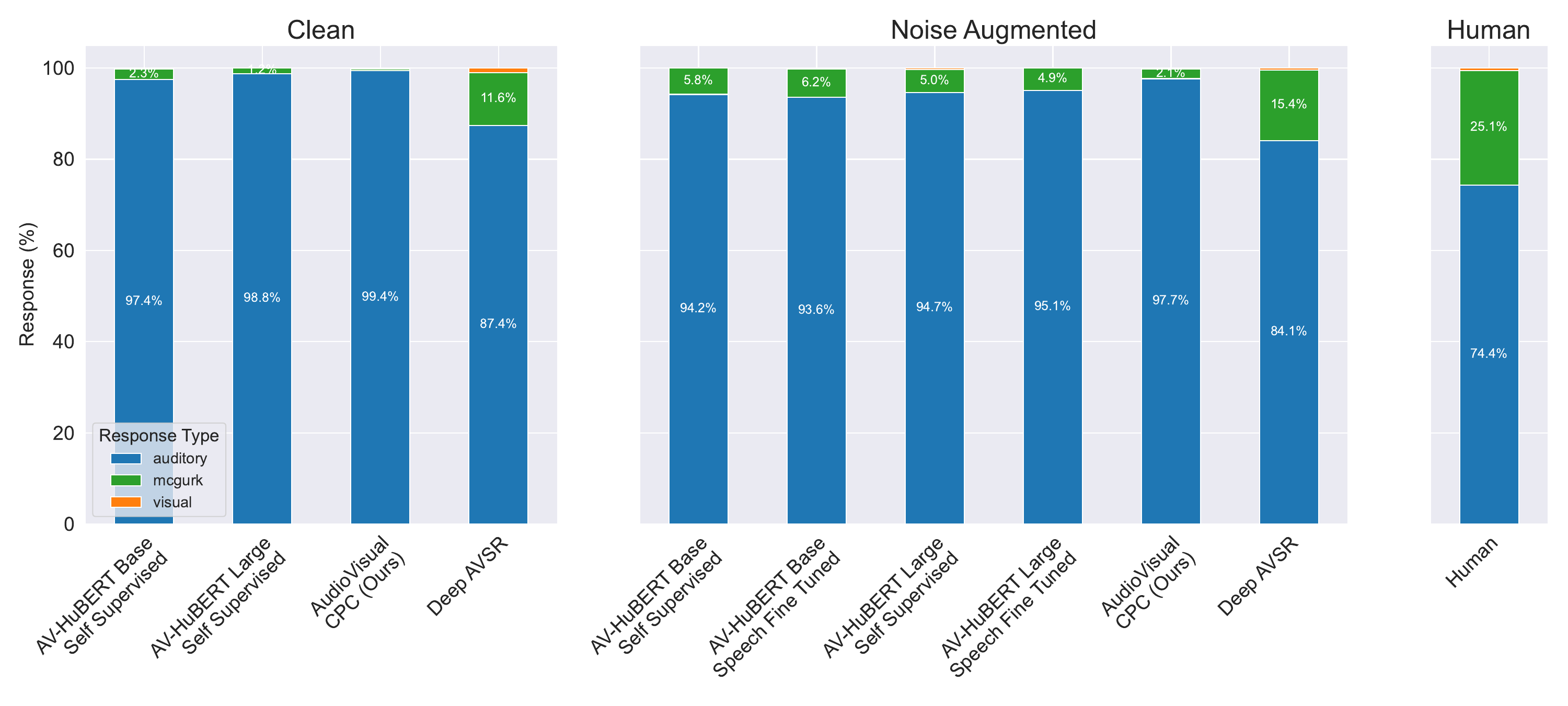}
    \caption{\textbf{Audio Only Control Stimuli Classification Results.} Networks were provided with auditory input and zeros for visual input, and human participants listened to speech and were presented with a blank screen. Networks primarily responded with auditory option and humans selected the auditory and McGurk options  25.1\% and 74.4\% of the time respectively.}
    \label{figure:knn_classification_audio}
\end{figure*}

Figure \ref{figure:knn_classification_audio}  shows the results of the ANNs and humans on audio-only trials in which the humans were presented audio and a blank screen and the ANNs were given zeros as input to their respective visual encoders. This demonstrates each networks ability to classify the audio of the McGurk stimuli set accurately in the absence of visual information.

\begin{figure}[!htbp]
  \centering
    \includegraphics[width=0.5\textwidth]{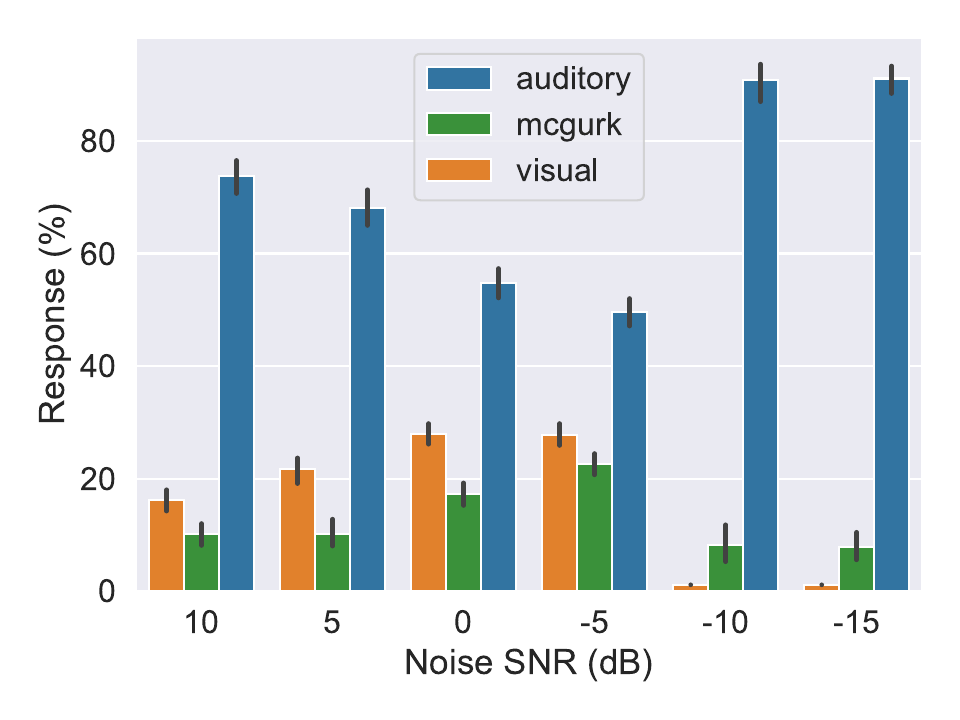}
    \caption{\textbf{AudioVisual CPC Network Performance Under Increased Levels of Training Noise.} Networks were trained with systematically increasing levels of babble noise in audio and were tested using clean audio and video. McGurk and visual responses increased with increased training noise, up to -5 dB SNR. At -10 dB and higher audiovisual integration broke and the network always selected the auditory response.
    Error bars indicate 95\% confidence intervals calculated using bootstrapping across the 10 cross-validation folds of the k-NN classifier.}
    \label{figure:noise}
\end{figure}

Figure \ref{figure:noise} shows the result of training the Audiovisual CPC network with increasing levels of noise. As the level of noise during training increased the percentage of mcgurk and visual responses also increased, up to -5 dB SNR. At -10 dB and higher, however, audiovisual integration broke and the network always selected the auditory response. This is despite the fact that the accuracy remained high at all levels of noise, as shown in Table \ref{tab:snr_cross_val_accuracy}.

\begin{figure*}[!htbp]
  \centering
    \includegraphics[width=1.0\textwidth]{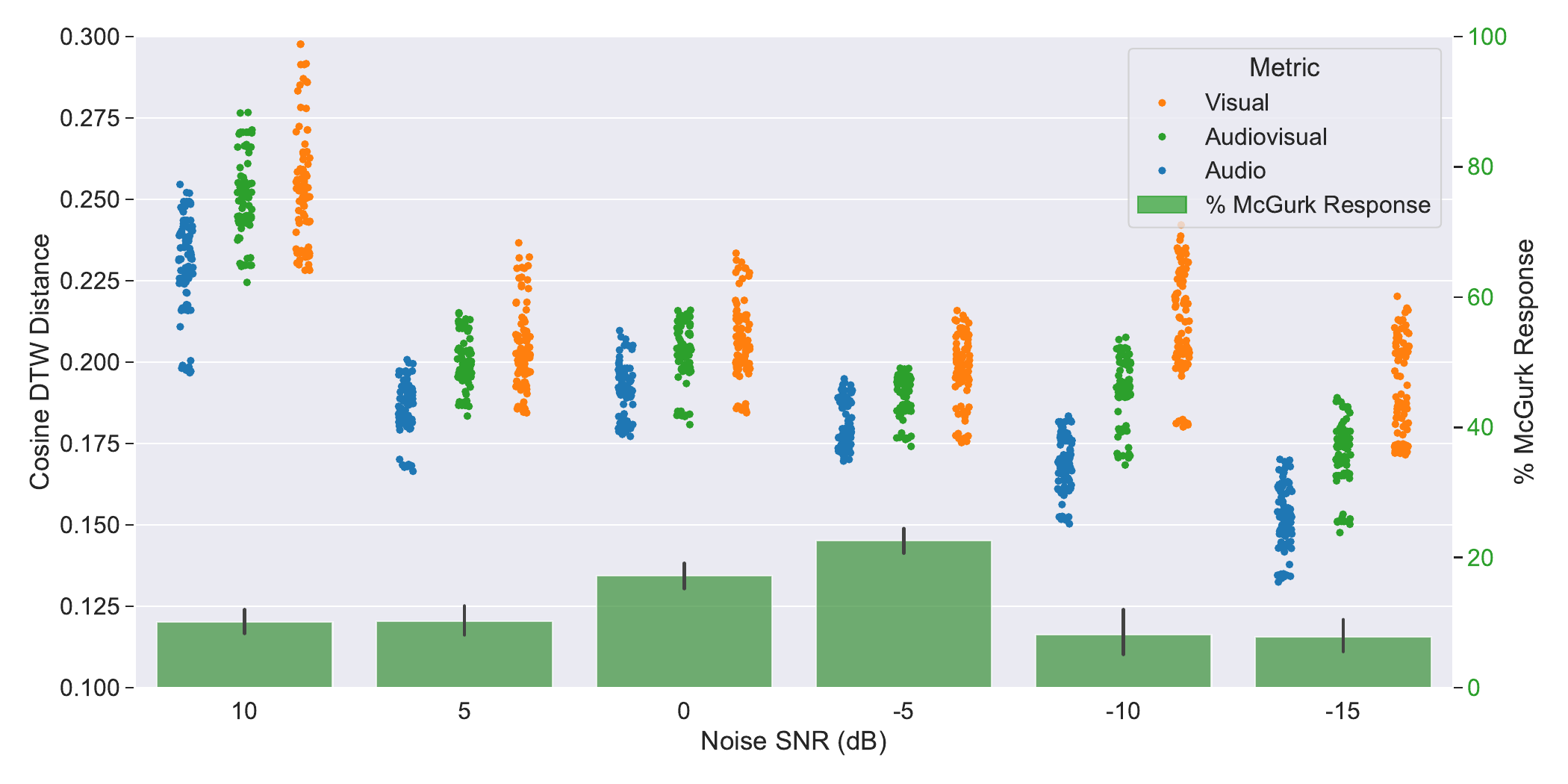}
    \caption{\textbf{AudioVisual CPC Embedding Distance Stripplot under Increased Levels of Training Noise.} Distances are the average cosine-DTW distance from each incongruent McGurk stimuli embedding to all word embeddings corresponding to the auditory option, visual option, and audiovisual option within each word pairing. The \% McGurk response from Figure \ref{figure:noise} is replotted for reference.}
    \label{figure:noise_distance}
\end{figure*}

Figure \ref{figure:noise_distance} further illuminates the effect of increased training noise by showing that the relative distance of the incongruent embedding to the auditory and audiovisual options decreases relative to the visual option. At high levels of noise (-10 dB SNR, -15 dB SNR), the auditory option is closest, followed by the audiovisual option and finally the visual option.

\begin{table}[!htbp]
\centering
\caption {\textbf{10-Fold Cross-Validation Accuracy on Congruent Stimuli Training Set at each Training SNR Level for AudioVisual CPC Network.} The shown 95\% confidence intervals are calculated using bootstrapping.}
\begin{tabular}{|l|l|l|}
\hline
SNR Level & Accuracy & 95\% CI \\ \hline
10 dB     & 97.78    & ± 2.04  \\ \hline
5 dB      & 97.04    & ± 2.22  \\ \hline
0 dB      & 95.93    & ± 2.41  \\ \hline
-5 dB     & 95.56    & ± 2.60  \\ \hline
-10 dB    & 96.67    & ± 2.22  \\ \hline
-15 dB    & 95.19    & ± 2.22  \\ \hline
\end{tabular}
\label{tab:snr_cross_val_accuracy}
\end{table}

\section{Discussion}

The McGurk Effect is a phenomenon that has been widely studied in humans and used as a proxy measure of audiovisual integration during speech perception. Previous research has shown the importance of visual information for the intelligibility of speech in noise \cite{sumby1954visual,bernstein2004auditory}, and the interaction between the presence of noise in audiovisual speech and stages of language development of audiovisual integration for speech perception \cite{hirst2018threshold}.  These observations refer to the presence of noise mixed with speech \textit{at the time of testing the listener}, not during the listener's period of learning and development. However, it suggests that noise present \textit{during development} might play a role in how audiovisual integration develops.  It is not possible to retroactively manipulate noise during human language learning.  Instead, in the present study we directly and systematically investigated the role of noise in the stimuli used to train audiovisual speech recognition neural networks in an effort to characterize the possible role of noise in human development. 

The specifics of how the McGurk Effect develops are not fully understood \cite{alsius2018forty}. The fact that some individuals do not develop the tendency to perceive the McGurk Effect and yet show no impairments in understanding congruent audiovisual information points to the need for a more thorough understanding of how the McGurk Effect emerges in development. Following from a recent framework that outlines how to ask developmental 'why' questions using ANNs \cite{kanwisher2023using}, we tested a sample of recent ANNs trained on audiovisual speech using audiovisually incongruent words designed to elicit a McGurk Effect. We compared networks trained with clean speech vs. noisy speech and found that training with noisy speech increased both visual-based responses and McGurk responses across all networks. We also found that systematically increasing the level of auditory noise during training of an ANN increased the tendency of the network to rely on visual input in general, and to exhibit the McGurk Effect in particular. However, this shift in reliance from auditory to visual or audiovisual input occurred only within a finite range of noise levels. Audiovisual integration failed to emerge when the network was trained with the two highest noise levels.  This shift away from audiovisual integration in favour of purely auditory perception occurred even though the network was still able to achieve high cross-validation scores during training. It is unclear why this shift occurred, but it suggests that high levels of noise pollution during early learning might have an important effect on audiovisual integration in humans as well.

\subsection{Comparison to Humans}

Although the networks showed a relative increase in McGurk responses, they did not select the McGurk option as often as human participants.  Furthermore, ANNs selected the visual option more often than human participants, who rarely select the visual option. This points to a difference in how transformer-based ANNs integrate visual information, even if they can achieve human levels of performance on speech recognition tasks. Furthermore, as shown in Figure \ref{figure:knn_classification_audio}, humans responded with the McGurk option at a higher rate than all ANNs when presented with only auditory stimuli. This is of note, as \cite{iqbal2023mcgurk} recently proposed that the McGurk effect is likely due to a default mechanism of the auditory system. They found a positive correlation with the default rate at which an individual responded with a given phoneme in an audio-only task and their response to McGurk stimuli using the same phoneme. This aligns with human responses from our audio-only condition and points to the need to interpret ANNs as models of human language with care.  Although they are likely to exhibit sensitivity to similar factors (e.g. the presence of noise during learning) they are not necessarily mechanistic models of human learning.   

\subsection{Insights about the McGurk Effect in Humans}

It is interesting to note that perception of the McGurk effect emerged \textit{untrained} in networks that were trained solely with \textit{self-supervised learning}. During this type of learning the network learns to represent inputs as meaningful feature embeddings without explicitly performing any supervised task, as the labels are derived from the data itself. This type of learning is somewhat more biologically plausible than \textit{supervised learning}, as humans are able to learn the phonetic features of speech in an unsupervised manner as early as in utero \cite{moon2013language}. The ability of self-supervised learning models to fuse auditory and visual speech into a representation akin to the McGurk percept points to the possibility that the McGurk effect might emerge through passive exposure to audiovisual speech throughout development.

The observation that the McGurk Effect occurs even when directly inspecting vector offsets in the embedding spaces of these ANNs lends some insight into the nature of the perceptual effect in humans.  In what sense is the McGurk Effect an illusion?  Perception is not always veridical. An illusion occurs when our perception does not accurately depict the physical world, and in this sense the McGurk Effect is indeed an illusion. In the McGurk Effect, listeners perceive a phoneme or word that was not present in the bottom-up sensory stream.  However, as revealed in Figure 4, suitably trained ANNs also map incongruent audiovisual input to vectors in their embedding spaces that are sometimes closer to the "illusory" McGurk word than to either auditory or visual words. Classifying encoded input vectors as the illusory "McGurk" word is therefore, in some sense, the best guess about the information being encoded by the input stage.  To the extent that such encoders might be models for early stages of audiovisual integration, this suggests that perceiving the McGurk Effect is only an illusion subsequent to encoding.  That is, perceiving the McGurk Effect (by humans or ANNs) is correct classification of incorrect encoding. 

The tendency to classify incongruent audiovisual pairs as McGurk words emerged in our CPC networks, and in Deep AVSR, despite the fact that these networks \textit{were never trained on incongruent stimuli}. Interestingly, AV-Hubert \textit{is trained with incongruent audiovisual speech} as a consequence of the masking approach used during training. This is due to the fact that AV-Hubert uses a novel masking-by-substitution approach in which masked visual segments are replaced with randomly selected visual segments from other training examples, which forces the network to identify the mismatch during self-supervised learning. As with human development, during training both Deep AVSR and our CPC networks were only ever presented with congruent audiovisual speech. They were never explicitly trained to exhibit the McGurk Effect.

A limitation of our lightweight Audiovisual CPC network is that although noisy audio increased the network's use of visual features, it did not fuse the audio and visual features into features aligned with the McGurk word at similar rates as AV-Hubert and Deep AVSR.  These are much larger networks with many transformer layers situated after an audiovisual fusion layer (12 for AV-Hubert Base, 24 for AV-Hubert Large, 6 for Deep AVSR). In contrast to this, our network consisted of a convolutional encoder and a single transformer layer as the context network. This suggests that deeper ANNs with multiple layers are likely required to effectively model the McGurk effect. This insight aligns with previous research into the McGurk effect using EEG. \cite{roa2015early}
 found that incongruent trials evoked a smaller N1 in the event-related potential than congruent trials, and that early and late beta power was suppressed in incongruent trials relative to congruent, respectively. The authors proposed a three-stage process for the McGurk effect in which the smaller N1 corresponds to enhanced audiovisual integration, the early beta power suppression corresponds to the detection of incongruent stimuli, and the late beta power suppression corresponds to the processing that integrates the incongruent stimuli into a more congruent perception. Future work could look for a similar pattern in ANNs by testing whether the internal activity of the ANN follows this pattern and, if multiple integration layers are required, whether they follow the pattern of early detection of incongruent stimuli and later processing to integrate the stimuli. 

\section{Conclusion}

This study demonstrated the use of ANNs as proxy models of one of the processes of speech and language development, namely the emergence of audiovisual integration.  This approach is informative in situations that call for systematic manipulation of one parameter (here, the presence of auditory noise during development) while holding other factors constant.  We showed that noise plays an important role in the learning process of audiovisual ANNs by pushing them to integrate visual information more prominently in the encoding of audiovisual speech.  

\section{Data Availability}
The audiovisual stimuli, k-NN classifier code, and human/neural network k-NN responses is available on OSF: \url{https://osf.io/3s5bq/?view_only=06ca5097cd4441d9b28654326a1519b1}. Code and model weights for Audiovisual CPC will be made available on github following publication.

\bibliographystyle{IEEEtran}
\bibliography{bibliography}

\end{document}